\documentclass[aps,prd,reprint,preprintnumbers,superscriptaddress,notitlepage,nofootinbib]{revtex4-1}

\usepackage[utf8]{inputenc}
\usepackage[english]{babel}
\usepackage{mathtools}
\usepackage{subdepth}
\usepackage{natbib}
\usepackage[bottom]{footmisc}
\usepackage{kotex}
\usepackage{graphicx}
\usepackage{float}
\usepackage{amssymb}
\usepackage{amsmath}
\usepackage{amsthm}
\usepackage{xcolor}
\usepackage{hyperref}
\usepackage{slashed}
\usepackage{kotex}
\usepackage[shortlabels]{enumitem}
\usepackage{microtype}
\usepackage{balance}
\usepackage[title]{appendix}
\usepackage[dvipsnames]{xcolor}
\usepackage{comment}
\usepackage{soul}

\begin{document}

\preprint{APCTP-Pre2026-014}

\title{
Quantum Clausius relation beyond de Sitter equilibrium
}

\author{Jinn-Ouk Gong}
\email[]{jgong@ewha.ac.kr}
\affiliation{Department of Science Education, Ewha Womans University, Seoul 03760, Korea}
\affiliation{Asia Pacific Center for Theoretical Physics, Pohang 37673, Korea}

\author{TaeHun Kim}
\email[]{gimthcha@kias.re.kr}
\affiliation{School of Physics, Korea Institute for Advanced Study, Seoul 02455, Korea}
\affiliation{International Center for Quantum-field Measurement Systems for Studies of the Universe and Particles, High Energy Accelerator Research Organization, Ibaraki 305-0801, Japan}
\affiliation{The Institute for Gravitation and the Cosmos, The Pennsylvania State University, University Park, Pennsylvania 16802, United States of America}

\author{Junghwan Lee}
\email[]{ghe126692@gmail.com}
\affiliation{Center for Theoretical Physics, Department of Physics and Astronomy, Seoul National University, Seoul 08826, Korea}

\author{Chang Sub Shin}
\email[]{csshin@cnu.ac.kr}
\affiliation{Department of Physics and Institute for Sciences of the Universe, Chungnam National University,
Daejeon 34134, Korea}
\affiliation{Particle Theory and Cosmology Group, Center for Theoretical Physics of the Universe,
Institute for Basic Science, Daejeon 34126, Korea}
\affiliation{School of Physics, Korea Institute for Advanced Study, Seoul 02455, Korea}


\begin{abstract}
We establish a direct quantum Clausius relation for conformal fields inside the apparent horizon of a quasi-de Sitter spacetime. We compute the net heat flow across the horizon from the renormalized stress-energy tensor, and independently determine the time evolution of the renormalized von Neumann entropy using the replica method. At leading order in the Hubble-flow expansion, the two calculations agree exactly, without using the gravitational equations of motion. Beyond leading order, we find a scheme-independent entropy balance that includes the corresponding Wald entropy. Our results provide a quantum field theoretic realization of dynamical horizon thermodynamics beyond exact de Sitter equilibrium.
\end{abstract}

\maketitle
\allowdisplaybreaks[4]
\maxdeadcycles=1000


\section{Introduction} 
\label{sec:intro}

The thermodynamic interpretation of causal horizons has revealed a profound connection among gravity, quantum field theory (QFT), and information. For quantum fields, tracing out degrees of freedom beyond a horizon gives rise to entanglement entropy with a universal ultraviolet (UV) structure. Its leading area divergence is naturally incorporated into the renormalization of Newton's constant in gravitational theories~\cite{Srednicki:1993im,Callan:1994py,Fursaev:1995ef,Jacobson:1994iw,Dvali:2008jb}. This area-law behavior applies equally to both black hole and cosmological horizons, while subleading terms encode information about the quantum field content. In even spacetime dimensions, in particular, a logarithmic contribution appears. For de Sitter (dS) space, this contribution has been obtained from both entanglement and thermal descriptions, providing a sharp QFT connection between horizon entanglement and the thermality of the dS static patch~\cite{Casini:2011kv,Eling:2013aqa}.

However, a dynamical realization of this connection remains less understood. In exact dS space, the dS-invariant vacuum is stationary and admits no net energy flow across the horizon, so the Clausius relation $\delta Q = T dS$ becomes trivial. A nonvanishing flux can instead arise from a state that breaks the dS isometries, such as an Unruh-de Sitter state~\cite{Aalsma:2019rpt,Gong:2020mbn}, while keeping the geometry in exact dS. At the opposite end, the thermodynamics of dynamical apparent horizons in the Friedmann-Lema\^itre-Robertson-Walker (FLRW) metric has been extensively studied by connecting the Friedmann equations with first-law or Clausius-like horizon relations~\cite{Frolov:2002va, Cai:2005ra, Akbar:2006kj, Akbar:2006mq, Cai:2006rs, Lidsey:2008zq}. In such constructions, however, the thermodynamic relation remains tied to the gravitational equations of motion. What has been missing is a direct QFT demonstration that the entropy of quantum fields inside an evolving cosmological horizon changes precisely as dictated by the heat flow across it, although leading-order demonstrations have been made in local causal diamonds~\cite{Alonso-Serrano:2020pcz} and stretched light cones~\cite{Alonso-Serrano:2025lbo}, much smaller than the curvature scale.

In this work, we provide such a demonstration for conformal fields inside the apparent horizon of a prescribed quasi-dS spacetime. The slow evolution of the Hubble rate provides a controlled departure from exact dS equilibrium, while conformal fields with no particle production provide clean probes. We compute the heat flow across the apparent horizon from the renormalized stress-energy tensor and the energy-supply relation, and independently determine the change in the renormalized von Neumann entropy using the replica method. We find that the two calculations exactly agree at leading order in the Hubble-flow expansion, providing a direct, nonvanishing quantum Clausius relation for an evolving cosmological horizon. Beyond leading order, the local $R^2$ sector of the gravitational effective action enters through its Wald entropy, and the entropy balance leaves a residual. However, the residual is not sign definite and cannot in general be identified with positive microscopic entropy production.

Crucially, no gravitational equation of motion is used to establish this quantum Clausius relation. Our result thus directly connects two previously distinct viewpoints -- quantum entanglement at cosmological horizons and the thermodynamic formulation of gravity in the FLRW metric. It is also a concrete step toward a QFT description of horizon thermodynamics in realistic, time-dependent cosmological settings.
Furthermore, the difference in algebraic types associated with exact dS~\cite{Chandrasekaran:2022cip} and quasi-dS~\cite{Seo:2022pqj} may imply a connection between the quantum Clausius relation and the algebraic structure of horizon observables. 

Throughout the article, we use the ``$+++$'' sign convention in Ref.~\cite{Misner:1973prb}, and the natural unit system of $c = \hbar = k_B = 1$. Roman indices $a, b, \cdots$ are used for the four-dimensional (4D) spacetime, and Greek indices $\alpha, \beta, \cdots$ are used for the 2D space appearing in Sec.~\ref{sec:entFlow}. The letter $S$ is used for the entropy, and we instead use $\mathcal{S}$ for the classical action.

\section{Entropy flow from the conformal anomaly} \label{sec:entFlow}

Whenever energy crosses the boundary of a thermodynamic system as heat, it carries an associated entropy flow. The entropy balance is written as
\begin{equation}
    \dot{S} = \frac{d_e S}{dt} + \frac{d_i S}{dt} \, , \label{eq:entropyBalance}
\end{equation}
where $d_e S$ is the external entropy flow transferred into the system and $d_i S$ is a residual between the total entropy change and the entropy flow. In conventional nonequilibrium thermodynamics, $d_i S$ is interpreted as internal entropy production which obeys $d_iS\geq0$~\cite{kondepudi2014modern}. For inward heat flow $\delta Q$ through a boundary at temperature $T$, $d_e S = \delta Q / T$.

We apply Eq.~\eqref{eq:entropyBalance} to the region inside the apparent horizon of a prescribed quasi-dS spacetime. We describe the geometry by a spatially flat FLRW metric $ds^2 = - dt^2 + a^2(t) \delta_{ij} dx^idx^j$ with Hubble rate $H=\dot a/a$, and the apparent horizon is at a comoving radius of $r_A=(aH)^{-1}$. Exact dS space in the dS-invariant vacuum is stationary, so both the entropy and heat rates vanish. Quasi-dS evolution supplies a controlled departure from that equilibrium while preserving spatial homogeneity and isotropy.

For a homogeneous and isotropic source with $T^a{}_b={\rm diag}(-\rho,p,p,p)$, the energy transferred across a spherical surface is described by the energy-supply one-form of the unified first law $\Psi_\alpha=\frac{1}{2}(\rho+p)(-Har,a)$, in the 2D orbit space of time and surface radius~\cite{Hayward:1997jp,Hayward:1998ee}. If a 2D vector $\zeta^\alpha$ specifies the displacement of the surface, the inward heat flow is $\delta Q=A\Psi_\alpha\zeta^\alpha$, where $A$ is the surface area; we set the inward direction positive. The two standard apparent horizon prescriptions pair different displacements with different temperatures~\cite{Cai:2005ra, Cai:2006rs}, but they agree in the entropy flow as (see App.~\ref{app:tempConf}) 
\begin{equation}
    \frac{1}{T}\frac{\delta Q}{dt}
    =-\frac{8\pi^2}{H^3}(\rho+p) \, . \label{eq:entropyflow}
\end{equation}

We now apply Eq.~\eqref{eq:entropyflow} to free conformal spectator fields, collectively denoted by $X$: a real conformally coupled scalar $\varphi$, a massless Dirac fermion $\psi$, and a massless vector field $V$. Their bulk equations of motion reduce to those in Minkowski space after conformal rescaling, so the conformal vacuum contains no particle production. Any nonzero $\rho+p$, and hence the heat and the entropy flow below, is therefore a curvature-induced vacuum-polarization effect rather than a flux of produced particles. This makes conformal fields particularly clean probes for dynamical horizon entropy balance.

Although the classical stress tensor is traceless, renormalization on a curved background generates the Weyl anomaly~\cite{parker1978aspects,Deser:1993yx},
\begin{align}
    \langle T^a{}_a\rangle^{(s)}_{\rm rn} ={}& \frac{1}{2880\pi^2}\left[\frac{\alpha}{2} \left(R_{abcd}R^{abcd}-4R_{ab}R^{ab}+R^2\right) \right. \nonumber \\
    & \qquad \quad \ \, \left. \vphantom{\frac{\alpha}{2}} + \beta^{(s)} \Box R + \gamma C_{abcd} C^{abcd} \right] \, ,
    \label{eq:TrAnomal}
\end{align}
where $\Box\equiv\nabla^a\nabla_a$, the subscript ``${\rm rn}$'' denotes a renormalized quantity, and the superscript $(s)$ labels the renormalization scheme. Coefficients $\alpha$ and $\gamma$ are universal for a given field content, whereas $\beta^{(s)}$ depends on the finite local prescription of renormalization. The three coefficients are determined by explicit calculations, such as $\zeta$-function regularization giving $(\alpha_\varphi,\beta_\varphi^{(s)},\gamma_\varphi)=(-1,1,3/2)$, $(\alpha_\psi,\beta_\psi^{(s)},\gamma_\psi)=(-11,6,9)$, and $(\alpha_V,\beta_V^{(s)},\gamma_V)=(-62,-18,18)$, while dimensional regularization changes the vector coefficient to $\beta_V^{(s)}=12$~\cite{parker1978aspects}.

In the flat FLRW metric, $C_{abcd}=0$ and $R_{abcd}R^{abcd}=2R_{ab}R^{ab}-R^2/3$. The trace and the conservation equation $\nabla_a\langle T^{ab}\rangle_{\rm rn}=0$ then determine the two independent components $\rho_{\rm rn}^{(s)}$ and $p_{\rm rn}^{(s)}$~\cite{Parker:2009uva,Maranon-Gonzalez:2024hbj}; explicit expressions are given in App.~\ref{app:derivRels}. For the coefficients, adiabatic subtraction agrees with $\zeta$-function regularization~\cite{Maranon-Gonzalez:2023efu,Maranon-Gonzalez:2024hbj}. Substituting the renormalized stress tensor into Eq.~\eqref{eq:entropyflow} gives the exact field contribution as
\begin{equation}
    \frac{1}{T}\frac{\delta Q_X^{(s)}}{dt} = \frac{\alpha}{90} H \epsilon + \beta^{(s)} H f(\epsilon) \, , \label{eq:entropyflowX}
\end{equation}
where
\begin{equation}
    f(\epsilon) \equiv \frac{\epsilon^2}{15} - \frac{\epsilon\epsilon_2}{60} - \frac{\epsilon^3}{30} + \frac{7\epsilon^2\epsilon_2}{180} - \frac{\epsilon\epsilon_2^2}{180} - \frac{\epsilon\epsilon_2\epsilon_3}{180} \, , \label{eq:fbeta}
\end{equation}
and $\epsilon \equiv -\dot H/H^2=\epsilon_1$ and $\epsilon_{n+1} \equiv \dot\epsilon_n/(H\epsilon_n)$ are Hubble-flow parameters. Equation~\eqref{eq:entropyflowX} involves no Hubble-flow truncation and vanishes in exact dS space. If all Hubble-flow parameters are counted as $\epsilon_i\ll1$ with no additional hierarchy~\cite{Choe:2004zg}, the universal $\alpha$ term begins at the first order, whereas the scheme-dependent $\beta^{(s)}$ term begins at the second order. This separation is especially transparent in the Hubble-flow parameterization; see App.~\ref{app:derivRels}.

The anomalous stress tensor has been widely studied in cosmology~\cite{Duff:1993wm}, including its relation to logarithmically corrected horizon thermodynamics associated with Friedmann equations~\cite{Lidsey:2008zq}. In the present work, we ask whether the entropy flow and the time dependence of the renormalized field entropy obey a Clausius relation on a prescribed background, without invoking gravitational equations of motion.

\section{Physical entropy flow}

The scheme dependence of $\beta^{(s)}$ is not a physical ambiguity of the complete heat flow. It reflects the freedom to transfer a finite local $R^2$ term between the matter and the local curvature sectors of the effective action~\cite{Deser:1993yx,Boulanger:2007ab,Shapiro:2008sf}. Keeping only the terms relevant here, the bare semiclassical action is $\mathcal{S} = \int d^4x\sqrt{-g} \,(\mathcal{L}_X + R/16\pi G^b + C^b_{R^2} R^2 +\cdots )$, and, after integrating out $X$ in a scheme $s$, the effective action is
\begin{equation}
    \Gamma_{\rm eff}
    =\int d^4x\sqrt{-g}
    \left(\frac{R}{16\pi G}+C^{(s)}_{R^2}R^2\right)
    +\Gamma_X^{(s)}[g]+\cdots \, .
    \label{eq:Gamma_eff}
\end{equation}
A finite reshuffling $\Gamma_X^{(s)} \rightarrow \Gamma_X^{(s)} + c \int d^4 x \sqrt{-g} \, R^2$ changes  $\beta^{(s)} \rightarrow \beta^{(s)} - 34560 \pi^2 c$ and $C_{R^2}^{(s)} \rightarrow C_{R^2}^{(s)} - c$, while leaving $\Gamma_{\rm eff}$ unchanged~\cite{parker1978aspects}. The matched combination
\begin{equation}
    C_{R^2}\equiv C_{R^2}^{(s)}
    -\frac{\beta^{(s)}}{34560\pi^2}
    \label{eq:CR2matched}
\end{equation}
is therefore independent of the local matter-curvature split. $C_{R^2}$ is not an accidental scheme-invariant combination, but is the coefficient of the complete local $R^2$ term in the renormalized total one-particle irreducible (1PI) functional, at a fixed curvature basis and low-energy matching condition; see Apps.~\ref{app:CR2matching} and \ref{app:scheme}. It therefore fixes the strength of the $R^2$ interaction in the corresponding $f(R)$ theory, including the Wald entropy in Sec.~\ref{sec:genEntropy}.

This scheme-invariant combination can also be read directly from the metric variation~\cite{Parker:1993dk},
\begin{equation}
    \delta\Gamma_{\rm eff}
    =\int d^4x\sqrt{-g}
    \left[
    \frac{G_{ab}}{16\pi G}
    +C_{R^2}^{(s)}H_{ab}^{(1)}
    -\frac{1}{2}(T_{ab})_{\rm rn}^{(s)}
    \right]\delta g^{ab},
    \label{eq:Geff-vary}
\end{equation}
where
$H_{ab}^{(1)}
=2RR_{ab}-\frac{1}{2}g_{ab}R^2
+2(g_{ab}\Box-\nabla_a\nabla_b)R$.
The scheme-dependent pieces combine as
\begin{equation}
    g^{ab}\left[(T_{ab})_{\rm rn}^{(s)}
    -2C_{R^2}^{(s)}H_{ab}^{(1)}\right]
    =-12C_{R^2}\Box R+\cdots \, ,
    \label{eq:non-univ-coeff}
\end{equation}
which is invariant under the finite reshuffling above.

Accordingly, at the level of the effective action, the $R^2$ terms from the matter and the gravity sectors should be combined to give the matched scheme-invariant $R^2$ contribution. Then the total effective action and the entropy flow are split into field and local pieces,
\begin{equation}
    \frac{d_e S}{dt} = \frac{1}{T} \left( \frac{\delta Q_{X}}{dt} + \frac{\delta Q_{R^2}}{dt} \right) \, ,  \label{eq:entropyflowXphysical}
\end{equation}
where
\begin{subequations}
\begin{align}
    \frac{1}{T} \frac{\delta Q_{X}}{dt}
    &= \frac{\alpha}{90}H \epsilon \, , \label{eq:entropyflowXphysicalX} \\
    \frac{1}{T} \frac{\delta Q_{R^2}}{dt} &= -34560\pi^2C_{R^2}Hf(\epsilon) \, . \label{eq:entropyflowXphysicalRsq}
\end{align} \label{eq:entropyflowXphysicaldecomp}%
\end{subequations}
In this decomposition, $\delta Q_X$ denotes the retained field contribution, while $\delta Q_{R^2}$ is now fully associated with the matched local $R^2$ term and begins only at second order in Hubble-flow expansion. When multiple fields are present, $\alpha$ in Eq.~\eqref{eq:entropyflowXphysicaldecomp} becomes the sum of $\alpha$ coefficients of all fields. We next compute the entropy side, independently of the stress tensor.

\section{von Neumann entropy of conformal fields}\label{sec:EE-conf}

For an entangling surface $\Sigma$, the replica method gives the von Neumann entropy~\cite{Callan:1994py,Solodukhin:2011gn}
\begin{equation}
    S_{\rm vN}
    =\left.\left(n\frac{\partial}{\partial n}-1\right)W(n)
    \right|_{n=1} \, ,
    \label{eq:SvN-def}
\end{equation}
where $W(n)$ is the effective action on the $n$-fold geometry around $\Sigma$. Writing
$W(n)=nW_{\rm bulk}+(1-n)W_\Sigma+\mathcal{O}[(1-n)^2]$ gives $S_{\rm vN}=-W_\Sigma$, and with a UV cutoff $\delta$, its UV structure is $S_{\rm vN} = C_X A_\Sigma / \delta^2 - W_\Sigma^{\rm log}\log\delta + S_{\rm finite}$, where the area divergence renormalizes Newton's constant and the logarithmic divergence renormalizes the local curvature-squared couplings~\cite{Fursaev:1994ea, Solodukhin:2011gn, Donnelly:2012st, Bousso:2015mna}. Hereafter, $S_{\rm vN}$ denotes the retained field entropy after the entropy contribution associated with the local $R^2$ term has been separated  off~\cite{Fursaev:1995ef,Cooperman:2013iqr}, consistently with the separation in Eqs.~\eqref{eq:entropyflowXphysicaldecomp}. The contact term from $\xi R \varphi^2$ is similarly assigned to the gravity side, in which its leading area divergence contributes to the renormalization of Newton's constant~\cite{Donnelly:2012st,Donnelly:2015hxa,Bousso:2015mna}.

The universal coefficient $W_\Sigma^{\rm log}$ is evaluated explicitly in App.~\ref{app:logCoeffFLRW}. For a spherical surface in a general flat FLRW background, all $\dot H$ contributions cancel and one obtains $ W_\Sigma^{\log,\varphi}=-1/90$, $W_\Sigma^{\log,\psi}=-11/90$, and $W_\Sigma^{\log,V}=-62/90$, identical to the exact dS values~\cite{Casini:2011kv, Eling:2013aqa}; the employed heat kernel method in the geometric replica construction includes edge-mode contribution for a vector field. The trace enters the entropy by
\begin{equation}
    \delta\frac{\partial S}{\partial\delta}
    =\left. \left(n\partial_n -1\right)
    \left\langle\int\sqrt{-g}\,
    T^a{}_a\right\rangle_{M_n}\right|_{n=1}  \, ,
    \label{eq:deltaS-Tranomal}
\end{equation}
so the surface coefficient is related to the bulk anomaly by
$\alpha=90W_\Sigma^{\rm log}$~\cite{Callan:1994py,Ryu:2006ef}. Our FLRW calculation thus verifies that $W_\Sigma^{\rm log}$ acquires no additional explicit dependence on $H(t)$ at the evolving apparent horizon. 

After the local UV terms are renormalized, the same coefficient $\alpha$ controls the time dependence of the remaining field entropy. Under a Weyl transformation of  $g_{ab}\rightarrow e^{2\sigma}g_{ab}$, the effective action changes by
\begin{equation}
    \delta_\sigma\Gamma_X^{(s)}
    =\int_M d^4x\,
    \delta_\sigma g_{ab}
    \frac{\delta\Gamma_X^{(s)}}{\delta g_{ab}}
    =\int_M d^4x\sqrt{-g}\,
    \langle T^a{}_a\rangle_{\rm rn}^{(s)}\sigma \, ,
    \label{eq:dW-Tanomal}
\end{equation}
and the corresponding change of the renormalized entropy at fixed coordinate surface is~\cite{Solodukhin:2008dh,Komargodski:2011vj,Solodukhin:2013yha}
\begin{align}
    &S_{\rm vN}^{\rm rn}(e^{2\sigma}g_{ab})
    -S_{\rm vN}^{\rm rn}(g_{ab}) \nonumber\\
    &\quad
    =\frac{1}{720\pi}
    \int_\Sigma\sqrt{\gamma}\,
    \left\{
    \alpha\left[\sigma R_\Sigma
    +(D_\Sigma\sigma)^2\right]
    +2\gamma\sigma K_\Sigma
    \right\} \, .
    \label{eq:fin-Weyl-entropy}
\end{align}
By taking $\sigma=\log a(t)$ and $g_{ab}=\eta_{ab}$ in conformal coordinates, $e^{2\sigma} g_{ab}$ becomes our flat FLRW metric and hence $S^{\rm rn}_{\rm vN} = S^{\rm rn}_{\rm vN} (a^2 g_{ab})$ here. The Weyl factor is constant on each horizon sphere so $D_\Sigma\sigma=0$, while $\int_\Sigma\sqrt{\gamma}\,R_\Sigma=8\pi$ and
$K_\Sigma=0$ for a sphere in a conformally flat background. These give (see App.~\ref{app:logCoeffFLRW}) 
\begin{equation}
    S_{\rm vN}^{\rm rn} - S_{\rm vN}^{\rm rn}(\eta_{ab}) = \frac{\alpha}{90} \log a(t) \, . \label{eq:SvNWeylTr}
\end{equation}

Since the apparent horizon has a time-dependent comoving radius, $S_{\rm vN}^{\rm rn}(\eta_{ab})$ must also be evaluated at
$r=(aH)^{-1}$. For a sphere of radius $r$, $S_{\rm vN}^{\rm rn}(\eta_{ab}) = (\alpha/90)\log(\mu r)+C_0$, where $\mu$ is a fixed renormalization scale and $C_0$ is independent of $r$~\cite{Solodukhin:2008dh}. Combination with Eq.~\eqref{eq:SvNWeylTr} gives $S_{\rm vN}^{\rm rn}=(\alpha/90)\log(\mu/H)+C_0$, and therefore
\begin{equation}
    \dot S_{\rm vN}^{\rm rn}
    =\frac{\alpha}{90}H\epsilon \, .
    \label{eq:dSvNdt}
\end{equation}
Eq.~\eqref{eq:dSvNdt} is a full-order expression in Hubble-flow parameters. We give a canonical density-matrix derivation of Eq.~\eqref{eq:dSvNdt} in App.~\ref{app:denmat}, which shows that the regulated entanglement spectrum in the FLRW metric is related to that in Minkowski space by a local unitary transformation.

Entanglement entropy in time-dependent FLRW metric has also been studied holographically~\cite{Giataganas:2021cwg}. In the present work, Eq.~\eqref{eq:dSvNdt} exactly matches the retained field contribution to the independently computed entropy flow in Eq.~\eqref{eq:entropyflowXphysicalX}. The equality is not automatic merely because the same anomaly coefficient appears on both sides, as the former follows from the replica method, whereas the latter follows from the conserved Lorentzian stress tensor and the horizon energy-supply relation. The geometrical factors also had to be properly considered to reach the agreement.

\section{Generalized entropy and the higher-order residual} \label{sec:genEntropy}

We now include the gravitational contribution on the entropy side. After the UV divergences of the field entropy are absorbed into the renormalized gravitational couplings, the generalized entropy can be decomposed into
\begin{equation}
    S_{\rm gen}^{\rm rn}
    =\frac{A}{4G}
    +S_{\rm grav}^{\rm rn}
    +S_{\rm vN}^{\rm rn} \, .
    \label{eq:Sgenrn}
\end{equation}
Here, $S_{\rm grav}^{\rm rn}$ denotes the entropy of local higher-curvature operators~\cite{Wald:1993nt, Jacobson:1993vj, Nelson:1994na, Iyer:1994ys, Fursaev:1995ef}. We treat the quasi-dS background as prescribed and the fields $X$ as spectators. Hence the Einstein-Hilbert area entropy is paired with an unknown background inflaton that supports this background and is therefore not included in Eq.~\eqref{eq:entropyflowXphysical}. The relevant entropy rate is instead $\dot S_{\rm vN}^{\rm rn}+\dot S_{W,R^2}$, where $S_{W, R^2}$ is the Wald entropy of the local $C_{R^2}R^2$ action term.

The Wald functional gives the horizon entropy contribution
\begin{equation}
    S_{W,R^2} = 8\pi C_{R^2}\int_\Sigma dA\,R = 192\pi^2C_{R^2}(2-\epsilon) \, ,
    \label{eq:SWR2}
\end{equation}
where $A=4\pi/H^2$ and $R=6(\dot H+2H^2)$. Its time derivative is thus
\begin{equation}
    \dot S_{W,R^2} = -192\pi^2C_{R^2}H\epsilon\epsilon_2 \, . \label{eq:dSWR2dt}
\end{equation}
For $f(R)$ interactions, no extrinsic curvature correction beyond the Wald expression is required~\cite{Dong:2013qoa, Camps:2013zua, Bousso:2015mna}. 

Substituting Eqs.~\eqref{eq:entropyflowXphysical}, \eqref{eq:dSvNdt}, and \eqref{eq:dSWR2dt} into Eq.~\eqref{eq:entropyBalance} gives the relevant entropy balance
\begin{equation}
    \dot S_{\rm vN}^{\rm rn} + \dot S_{W,R^2} = \frac{1}{T}\left(\frac{\delta Q_{X}}{dt} + \frac{\delta Q_{R^2}}{dt}\right) + \frac{d_iS}{dt} \, , \label{eq:entropyBalanceExpand}
\end{equation}
with the residual
\begin{align}
    \frac{d_i S}{dt} ={}& 34560\pi^2 C_{R^2} H \nonumber \\
    &\times \left( \frac{\epsilon^2}{15} - \frac{\epsilon \epsilon_2}{45} - \frac{\epsilon^3}{30}  + \frac{7 \epsilon^2 \epsilon_2}{180} - \frac{\epsilon \epsilon_2^2}{180} - \frac{\epsilon \epsilon_2 \epsilon_3}{180} \right) \, . \label{eq:diSdt}
\end{align}
This expression is exact in the flat FLRW metric. Defining $F\equiv1+32\pi GC_{R^2}R$, it admits a compact form
\begin{equation}
    \frac{d_iS}{dt} = \frac{\pi}{GH^3} \ddot F \, , \label{eq:diSdtCompact}
\end{equation}
whose derivation is given in App.~\ref{app:Einsteinframe}.

Both $\dot S_{W,R^2}$ and $d_iS/dt$ begin at second order in the Hubble-flow expansion, so the combined balance in Eq.~\eqref{eq:entropyBalanceExpand} reduces at leading order to
\begin{equation}
    \dot S_{\rm vN}^{\rm rn}
    = \frac{1}{T}\frac{\delta Q_X}{dt}
    = \frac{\alpha}{90}H\epsilon \, .
    \label{eq:leadingClausius}
\end{equation}
Importantly, this relation is exact in the retained field sector. This direct quantum Clausius relation follows from two independent QFT calculations on a prescribed background; no gravitational field equation is needed.

At the first nonvanishing order, the residual is
\begin{equation}
    \frac{d_iS}{dt}
    =768\pi^2C_{R^2}H\epsilon(3\epsilon-\epsilon_2)
    +\mathcal{O}(\epsilon_i^3) \, ,
    \label{eq:diSleading}
\end{equation}
which is not sign definite even for $C_{R^2}>0$. The residual therefore should not, by itself, be identified with a positive microscopic entropy production. Its local curvature structure agrees with the additional term in the FLRW apparent horizon nonequilibrium balance, obtained by rewriting the equations of motion of $f(R)$ gravity~\cite{Akbar:2006mq}. In the present work, this term accompanies the matched local curvature sector in the extended balance in Eq.~\eqref{eq:entropyBalanceExpand}, while the remaining field entropy continues to obey Eq.~\eqref{eq:leadingClausius}.

\section{Summary and discussion} \label{sec:discussion}

We have applied the entropy balance equation to free conformal fields inside the apparent horizon of a prescribed quasi-dS spacetime. The entropy flow was obtained from the renormalized stress tensor and the energy supply of the unified first law, while the entropy rate was obtained independently from the replica method. At leading Hubble-flow order, the two calculations give the same result as in Eq.~\eqref{eq:leadingClausius}, so the universal horizon heat flow has a direct quantum-state interpretation. For expanding backgrounds with $H > 0$ and $\epsilon > 0$, the renormalized entropy of the field inside the apparent horizon decreases while energy flows outward. This subareal decrease does not threaten a generalized second law. Once the background inflaton and Einstein-Hilbert area term are restored, the area contribution dominates parametrically, $|\dot S_{\rm vN}^{\rm rn}|/\dot S_{\rm area}=|\alpha|GH^2/(180\pi) \ll 1$, unless field multiplicity topples the hierarchy by $\alpha$.

Beyond leading order, the comparison cannot be made with the scheme-dependent field stress tensor alone. A finite change of the $\Box R$ term of the trace anomaly is inseparable from a compensating change of the local $R^2$ coupling. Matching these pieces in the complete effective action produces the scheme-independent coefficient $C_{R^2}$ and fixes both the entropy flow and the Wald entropy. Their mismatch is the residual in Eqs.~\eqref{eq:diSdt} and \eqref{eq:diSdtCompact}. Because this residual is not sign definite, it does not define a positive entropy production. 

Our approach separately treated the subareal entropy balance. If the background inflaton is included, its heat flow and the Einstein-Hilbert area entropy restore the usual leading entropy balance without changing the residual $d_i S / dt$~\cite{Frolov:2002va}. If all matched local curvature terms are assigned to the background gravitational sector, the remaining conformal field subsystem instead obeys the exact quantum Clausius relation Eq.~\eqref{eq:leadingClausius} without a separated $R^2$ residual. These are equivalent organizations of the same effective theory; the familiar nonequilibrium form of $f(R)$-gravity horizon thermodynamics is one such organization~\cite{Eling:2006aw,Akbar:2006mq}.

The same point can be geometrically viewed in an Einstein-frame description. The conformal metric redefinition trades the $R^2$ gravity for an explicit scalaron field and simultaneously changes the intrinsic apparent horizon and the time variable. The Jordan-frame residual is then decomposed into the difference between the entropy rates on the two horizon surfaces, the change in the inflaton entropy flow normalization, and the canonical scalaron heat flow; see App.~\ref{app:Einsteinframe}.

Our analysis is restricted to free conformal fields in the conformal vacuum on a spatially flat FLRW background, with the quasi-dS history treated as prescribed. We have not established a generalized second law for an apparent horizon, nor identified $d_i S$ with a positive microscopic entropy production. For massive or nonconformal fields, non-vacuum states, or interacting sectors, particle production and genuinely state-dependent evolution of the entanglement spectrum can arise. These systems provide the next test of how far the leading quantum Clausius relation persists, while also being more relevant to cosmological considerations. For an operator-algebraic perspective, the Type $\mathrm{II}_1$ static patch algebra of exact dS~\cite{Chandrasekaran:2022cip} and the proposed Type $\mathrm{II}_\infty$ description of inflationary quasi-dS~\cite{Seo:2022pqj} suggest that the quantum Clausius relation found here may admit an operator-algebraic interpretation. All these are left as future work.

\acknowledgments 

J.G. is supported in part by the National Research Foundation of Korea (NRF) grant RS-2024-00336507 and the Ewha Womans University Research Grant of 2026 1-2026-0486-001-1. 
T.H.K. is supported by KIAS Individual Grant PG095202 at Korea Institute for Advanced Study.
J.H.L. is supported by the NRF grants 2019R1C1C1010050 and RS-2024-00342093. 
C.S.S. is supported by Global-Learning \& Academic research institution for Master’s, PhD students, and Postdocs (G-LAMP) Program of the NRF grant RS-2025025442707, and by the NRF grant  RS-2026- 25498521. 
J.G. is also grateful to the Asia Pacific Center for Theoretical Physics for hospitality while this work was in progress.

\appendix

\begin{widetext}

\section{Temperature prescriptions and entropy flow at the FLRW apparent horizon} \label{app:tempConf}

For a dynamical apparent horizon, the infinitesimal displacement $\zeta^\alpha$ must specify the physical process being considered. Two standard choices are an isochoric process, in which the areal radius is instantaneously held at the horizon value~\cite{Cai:2005ra}, and a process that follows the moving apparent horizon~\cite{Cai:2006rs}. Since the apparent horizon of the flat FLRW metric is at $r_A=(aH)^{-1}$, the corresponding displacement vectors are
\begin{equation}
    \zeta_H^\alpha = \left(1,-\frac{1}{a}\right)dt   \quad \text{and} \quad \zeta_\kappa^\alpha = \left(1,-\frac{1}{a}(1-\epsilon)\right)dt \, . \label{eq:zetaUFL}
\end{equation}
Substituting these into $\delta Q=A\Psi_\alpha\zeta^\alpha$ gives
\begin{equation}
    \delta Q_H = -\frac{4\pi}{H^2}(\rho+p)\,dt \quad \text{and} \quad \delta Q_\kappa = -\frac{4\pi}{H^2}(\rho+p) \left(1-\frac{\epsilon}{2}\right)dt \, . \label{eq:deltaQ}
\end{equation}

The same factor distinguishes the two associated temperature prescriptions, often referred to as the temperature confusion~\cite{Tian:2014sca},
\begin{equation}
    T_H = \frac{H}{2\pi} \quad \text{and} \quad T_\kappa = \frac{|\kappa|}{2\pi} = \frac{H}{2\pi}\left(1-\frac{\epsilon}{2}\right) \, , \label{eq:T}
\end{equation}
where $\kappa=-H(1-\epsilon/2)$ is the surface gravity at the apparent horizon (quasi-dS spacetime with $\epsilon \ll 1$ does not flip the sign of $\kappa$). In Einstein gravity, each pair $(\delta Q_H,T_H)$ and $(\delta Q_\kappa,T_\kappa)$ satisfies the Clausius relation~\cite{Cai:2005ra,Cai:2006rs}. The temperature $T_H$ is obtained in an instantaneous horizon approximation to tunneling, whereas dynamical horizon treatments motivate $T_\kappa$~\cite{Parikh:1999mf,Cai:2008gw,Helou:2015yqa}. Our comparison uses the identical ratio $\delta Q/T$ obtained from the two consistently paired prescriptions, without requiring the temperatures to be operationally equivalent.

For either consistent pair, the entropy flow rate is identical,
\begin{align}
    \frac{1}{T}\frac{\delta Q}{dt}
    &\equiv \frac{1}{T_H}\frac{\delta Q_H}{dt} =\frac{1}{T_\kappa}\frac{\delta Q_\kappa}{dt} \nonumber\\
    &= -\frac{8\pi^2}{H^3}(\rho+p) \, . \label{eq:entropyflowAppendix}
\end{align}
This is Eq.~\eqref{eq:entropyflow} of the main text.

\section{Hubble-flow parameters and entropy flow expansion} \label{app:derivRels}

We collect the relations used to obtain Eqs.~\eqref{eq:entropyflowX} and \eqref{eq:fbeta}. Up to the fourth order in time derivative for the scale factor,
\begin{subequations}
\begin{align}
    \dot{a} &= aH \, , \\
    \ddot{a} &= a(\dot H+H^2) \, , \\
    \dddot{a} &= a(\ddot H+3H\dot H+H^3) \, , \\
    \ddddot{a} &= a(\dddot H+4H\ddot H+3\dot H^2+6H^2\dot H+H^4) \, .
\end{align} \label{eq:adots}%
\end{subequations}
The Hubble-flow parameters are defined by
\begin{equation}
    \epsilon_1=-\frac{\dot H}{H^2}\equiv\epsilon
    \quad \text{and} \quad
    \epsilon_{n+1}=\frac{\dot\epsilon_n}{H\epsilon_n} \, ,
    \label{eq:HubbleFlowDef}
\end{equation}
which gives
\begin{subequations}
\begin{align}
    \frac{\dot H}{H^2} &= -\epsilon \, , \\
    \frac{\ddot H}{H^3} &= 2\epsilon^2-\epsilon\epsilon_2 \, , \\
    \frac{\dddot H}{H^4}
    &=-\left(6\epsilon^3-7\epsilon^2\epsilon_2 +\epsilon\epsilon_2^2+\epsilon\epsilon_2\epsilon_3\right) \, .
\end{align} \label{eq:Hdots}%
\end{subequations}

In the flat FLRW metric, the anomaly in Eq.~\eqref{eq:TrAnomal} reduces to
\begin{equation}
    \langle T^a{}_a\rangle_{\rm rn}^{(s)}
    =\frac{1}{2880\pi^2}
    \left[
    -\alpha\left(R_{ab}R^{ab}-\frac{1}{3}R^2\right)
    +\beta^{(s)}\Box R
    \right] \, .
    \label{eq:TrAnomalFLRW}
\end{equation}
Together with $\nabla_a\langle T^{ab}\rangle_{\rm rn}^{(s)}=0$, this gives the renormalized energy density and pressure as~\cite{Parker:2009uva,Maranon-Gonzalez:2023efu,Maranon-Gonzalez:2024hbj}
\begin{align}
    \rho_{\rm rn}^{(s)}
    ={}&
    -\frac{(\alpha+3\beta^{(s)})\dot a^4}{960\pi^2a^4}
    +\frac{\beta^{(s)}\ddot a\dot a^2}{480\pi^2a^3}
    -\frac{\beta^{(s)}\ddot a^2}{960\pi^2a^2}
    +\frac{\beta^{(s)}\dddot a\dot a}{480\pi^2a^2} \, ,
    \label{eq:rhorn}\\
    p_{\rm rn}^{(s)}
    ={}&
    -\frac{(\alpha+3\beta^{(s)})\dot a^4}{2880\pi^2a^4}
    +\frac{(\alpha+3\beta^{(s)})\ddot a\dot a^2}{720\pi^2a^3}
    -\frac{\beta^{(s)}\ddot a^2}{960\pi^2a^2}
    -\frac{\beta^{(s)}\dddot a\dot a}{720\pi^2a^2}
    -\frac{\beta^{(s)}\ddddot a}{1440\pi^2a} \, .
    \label{eq:prn}
\end{align}
The integration constant has been set to zero, as appropriate to the conformal vacuum with no particle excitations. Their sum is
\begin{equation}
    \rho_{\rm rn}^{(s)}+p_{\rm rn}^{(s)}
    =
    -\frac{(\alpha+3\beta^{(s)})\dot a^4}{720\pi^2a^4}
    +\frac{(2\alpha+9\beta^{(s)})\ddot a\dot a^2}{1440\pi^2a^3}
    -\frac{\beta^{(s)}\ddot a^2}{480\pi^2a^2}
    +\frac{\beta^{(s)}\dddot a\dot a}{1440\pi^2a^2}
    -\frac{\beta^{(s)}\ddddot a}{1440\pi^2a} \, .
    \label{eq:rhorn+prn}
\end{equation}
Substituting Eq.~\eqref{eq:rhorn+prn} into Eq.~\eqref{eq:entropyflow} and using Eqs.~\eqref{eq:adots} and \eqref{eq:Hdots} gives Eqs.~\eqref{eq:entropyflowX} and \eqref{eq:fbeta}. In the form of Eq.~\eqref{eq:rhorn+prn}, the cancellation of the exact dS contribution can be anticipated, but the disappearance of $\beta^{(s)}$ at the first Hubble-flow order is not manifest. It becomes apparent only after the conversion to $\epsilon_i$, where all $\mathcal{O}(\epsilon)$ terms proportional to $\beta^{(s)}$ cancel. The matched rate of Eq.~\eqref{eq:entropyflowXphysical} inherits this separation because it contains the same function $f(\epsilon)$.

\section{Local $R^2$ coefficient in the renormalized 1PI effective action}
\label{app:CR2matching}

In this appendix, we identify the matched coefficient $C_{R^2}$ in Eq.~\eqref{eq:CR2matched} as the coefficient of the $R^2$ term of the complete renormalized 1PI effective action. In doing so, we decompose the four-derivative terms according to the spin structure, and identify the part that the $R^2$ term contributes, and then determine its coefficient. Throughout the discussion, we implement a convenient curvature basis of $(C_{abcd}C^{abcd},E_4,R^2,\Box R)$~\cite{Shapiro:2008sf,Decanini:2005eg}, and fix the renormalization scale and low-energy matching condition.

We first show that in the chosen curvature basis, the spin-zero part of the local four-derivative terms is carried solely by the $R^2$ term. To see this explicitly, expand the metric as $g_{ab}=\eta_{ab}+h_{ab}$. Since the background curvature vanishes, the quadratic part of the $R^2$ effective action term is
\begin{equation}
    \left. \int d^4x\sqrt{-g} \, R^2\right|_{h^2} = \int d^4 x \, \left( R^{(1)}[h] \right)^2 \, ,
\end{equation}
where $R^{(1)}[h] = \partial_a \partial_b h^{ab} - \Box h$ depends only on the gauge-invariant scalar component of the metric perturbation. Accordingly, this quadratic term is proportional to the spin-zero projector
\begin{equation}\label{eq:spin-0_project}
    P_{abcd}^{(0)}
    \equiv
    \frac{1}{3}
    \theta_{ab}\theta_{cd} 
    \quad \text{with} \quad
    \theta_{ab}
    \equiv
    \eta_{ab}
    -\frac{p_a p_b}{p^2} \, .
\end{equation}
On the other hand, other operators do not provide this spin-zero contribution~\cite{Stelle:1976gc,Alvarez:2018}. Thus, within this fixed curvature basis, the coefficient of the local four-derivative spin-zero kernel is precisely the coefficient multiplying $R^2$ which we discuss below.

We now define
\begin{equation}
    I_{R^2}[g] \equiv \int d^4x\sqrt{-g} \, R^2 \, . \label{eq:IR2functional}
\end{equation}
We write the local $R^2$ contribution to the effective action of a field $X$ as
\begin{equation}
    \left. \Gamma_X^{(s)} \right|_{\mathrm{local}, \, 0, \, \partial^4}
    =
    c_{X,R^2}^{(s)} I_{R^2}[g] \, ,
    \label{eq:GammaXlocalR2}
\end{equation}
where $c_{X,R^2}^{(s)}$ denotes the coefficient of the local $R^2$ term contained in the matter functional $\Gamma_X^{(s)}$, hence is distinct from the coefficient $C_{R^2}^{(s)}$ in the local gravitational sector of Eq.~\eqref{eq:Gamma_eff}. We remark that neither coefficient is separately independent of the renormalization prescription.

For the conformally coupled scalar, the presence of such a local term can be seen explicitly in the two-point function of the stress-tensor evaluated at one-loop order. At $\xi=1/6$, the nonlocal spin-zero contribution proportional to $p^4\log(-p^2/\mu^2)$ vanishes, whereas the finite spin-zero polynomial is local and can be represented by the $R^2$ term in the 1PI effective action~\cite{CasarinGodazgarNicolai:2018}. For fermions and vectors, the relation needed below follows more generally from the standard connection between the prescription-dependent $\Box R$ term in the trace anomaly and the freedom to add a finite local $R^2$ counterterm~\cite{AsoreyGorbarShapiro:2004}.

To determine the coefficient of this local term, consider an infinitesimal local Weyl transformation of the metric, $\delta_\sigma g_{ab}=2\sigma g_{ab}$, where the effective action transforms under Eq.~\eqref{eq:dW-Tanomal}. For the local functional $I_{R^2}[g]$ in Eq.~\eqref{eq:IR2functional}, the same transformation gives
\begin{equation}
    \delta_\sigma I_{R^2}[g] = -12\int d^4x\sqrt{-g}\,R\Box\sigma = -12\int d^4x\sqrt{-g}\,\sigma\Box R \, ,
    \label{eq:WeylR2app}
\end{equation}
where boundary terms have been omitted. By matching the integrands for arbitrary $\sigma$, it follows that the local term in Eq.~\eqref{eq:GammaXlocalR2} contributes
\begin{equation}
    \left.
    \left\langle T^a{}_a\right\rangle_{\rm rn}^{(s)}
    \right|_{\mathrm{local},\,0}
    =
    -12c_{X,R^2}^{(s)}\Box R \, .
    \label{eq:TraceFromLocalR2}
\end{equation}
Comparing this expression with the normalization of the $\Box R$ term in the trace anomaly in Eq.~\eqref{eq:TrAnomal},
\begin{equation}
    \left.
    \left\langle T^a{}_a\right\rangle_{\rm rn}^{(s)}
    \right|_{\Box R}
    =
    \frac{\beta^{(s)}}{2880\pi^2}\Box R \, ,
\end{equation}
we have
\begin{equation}
    c_{X,R^2}^{(s)} = - \frac{\beta^{(s)}}{34560\pi^2} \, . \label{eq:cXbeta}
\end{equation}
Thus, for a fixed renormalization prescription, the coefficient $c_{X,R^2}^{(s)}$ of the local $R^2$ term in the matter effective action corresponds to the coefficient $\beta^{(s)}$ of the $\Box R$ term in the trace anomaly. 

The complete local $R^2$ term in the renormalized total 1PI effective action is therefore  
\begin{equation}
    \left.
    \Gamma_{\rm eff}
    \right|_{\mathrm{local},\,R^2}
    =
    \left[
    C_{R^2}^{(s)}
    +
    c_{X,R^2}^{(s)}
    \right]
    I_{R^2}[g]
    =
    C_{R^2} I_{R^2}[g] \, ,
    \label{eq:GammaTotalLocalR2}
\end{equation}
where
\begin{equation}
    C_{R^2}
    =
    C_{R^2}^{(s)}
    +c_{X,R^2}^{(s)}
    =
    C_{R^2}^{(s)}
    -\frac{\beta^{(s)}}{34560\pi^2}
    \label{eq:CR2total1PI}
\end{equation}
as in Eq.~\eqref{eq:CR2matched}. For cases with multiple conformal fields, $c_{X,R^2}^{(s)}$ and $\beta^{(s)}$ should be the corresponding sums over the fields.

The same coefficient can be specified by a low-energy matching condition on the 1PI self energy of the metric. To make this explicit, we once again expand the metric as $g_{ab}=\eta_{ab}+h_{ab}$, and write the four-derivative spin-zero contribution as 
\begin{equation}
    \left.
    \Gamma_{\rm eff}^{(2)}
    \right|_0
    =
    \frac{1}{2}
    \int\frac{d^4p}{(2\pi)^4}\,
    h^{ab}(-p)p^4
    \Pi_{0,\rm tot}(p^2)
    P_{abcd}^{(0)}
    h^{cd}(p) \, .
    \label{eq:GammaSpinZeroMatching}
\end{equation}
At the quadratic order of $h_{ab}$, the local $R^2$ term in the momentum space gives
\begin{equation}
    \left.
    C_{R^2}I_{R^2}
    \right|_{h^2}
    =
    3C_{R^2}
    \int\frac{d^4p}{(2\pi)^4}\,
    h^{ab}(-p)p^4
    P_{abcd}^{(0)}
    h^{cd}(p) \, .
    \label{eq:R2SpinZeroMatching}
\end{equation}
Matching Eqs.~\eqref{eq:GammaSpinZeroMatching} and \eqref{eq:R2SpinZeroMatching} at the chosen scale gives
\begin{equation}
    \Pi_{0,\rm tot}^{\rm local}
    =
    6C_{R^2} \, .
    \label{eq:Pi0CR2}
\end{equation}
Thus, the matching specifies the local part of this form factor at the chosen renormalization scale. Any nonanalytic contributions such as logarithmic terms must be separated according to the same prescription, and higher-derivative terms must be treated separately. This condition yields the local four-derivative coefficient, not the total two-point function.

Once $C_{R^2}$ is shown to be the coefficient of the $R^2$ term of the total 1PI effective action by the matching at a scale, it should be invariant under a finite change of renormalization prescription that we consider. For example, if we vary the prescription by adding a finite local $R^2$ counterterm to the matter effective action, the renormalized gravitational coefficient must be shifted by compensating the same amount, leaving the action and hence the coefficient $C_{R^2}$ invariant.

The invariance established above is restricted to finite counterterm changes at fixed curvature basis and matching condition. It does not imply $C_{R^2}$ is invariant under redefinitions of the metric field. Such redefinitions can move contributions between curvature operators while preserving observables when considering the complete action consistently~\cite{Criado:2018sdb}.

As it is now clear that $C_{R^2}$ determines the $R^2$ part of the $f(R)$ gravity theory itself, the same coefficient also governs the corresponding entropy contribution. The corresponding Wald contribution is determined by $C_{R^2}$ with no additional extrinsic curvature term~\cite{Dong:2013qoa, Camps:2013zua, Bousso:2015mna} as in Eq.~\eqref{eq:SWR2}.

\section{Curvature-squared terms in flat FLRW} \label{app:scheme}

Appendix~\ref{app:CR2matching} identified $C_{R^2}$ as the coefficient of the complete local $R^2$ term in the renormalized effective action. Within the adopted semiclassical framework, our calculation concerns the local curvature terms associated with the 4D trace anomaly, and we therefore work consistently at curvature-squared order~\cite{Shapiro:2008sf}. We now explain why, at this order, it is sufficient to retain the $R^2$ term as a representative of the local curvature contribution to the time-dependent flat FLRW entropy balance. Higher-curvature operators and quantum-gravitational corrections from graviton loops lie outside the scope of the present work.

At four-derivative order, a convenient curvature basis of the renormalized gravitational effective action is formed by $E_4$, $C_{abcd}C^{abcd}$, $R^2$, and $\Box R$~\cite{Shapiro:2008sf,Decanini:2005eg}. On a spatially flat FLRW background, the Weyl tensor and hence the bulk metric variation generated by $C_{abcd}C^{abcd}$ vanish, and the corresponding surface contribution also vanishes for the spherical horizon, as follows from Eq.~\eqref{eq:Curv-vanish}. The integral of $E_4$ is topological, so its entropy contribution is constant under time evolution of a closed spherical surface. Finally, $\Box R$ is a total derivative action term which does not contribute to the metric variation in the bulk, and is also known to give no contribution to the entropy~\cite{Dong:2015zba}. Thus, in this curvature basis, only $R^2$ gives a nontrivial time dependence to the entropy balance.

A finite $R^2$ term changes the trace by a multiple of $\Box R$, as shown explicitly in Eq.~\eqref{eq:WeylR2app}. The prescription dependence of $\beta^{(s)}$ is therefore incorporated into the total coefficient $C_{R^2}$ of Eq.~\eqref{eq:CR2matched}. The same conclusion is obtained in any other curvature basis, provided that all coefficients are matched consistently. For example, one may use $(R^2, R_{ab}R^{ab}, R_{abcd} R^{abcd}, \Box R)$ instead of $(C_{abcd} C^{abcd}, E_4, R^2, \Box R)$. In the former basis, the bulk metric variations of the three quadratic curvature terms obey one linear relation following from the 4D Gauss--Bonnet identity, leaving only two independent components~\cite{Decanini:2005eg}. Moreover, the traces of all three variations are proportional to $\Box R$. Consequently, their finite local redefinitions project onto the same $\beta^{(s)}$ ambiguity in the trace anomaly.

For the entropy calculation, the $f(R)$ representative is particularly convenient. For a Lagrangian depending on the curvature through $R$ alone, the extrinsic curvature part of the generalized higher-curvature entropy functional vanishes, and the result reduces to the Wald functional~\cite{Dong:2013qoa, Camps:2013zua, Bousso:2015mna}. This justifies the use of Eqs.~\eqref{eq:SWR2} and \eqref{eq:dSWR2dt} and permits a direct comparison with the established nonequilibrium horizon thermodynamics of $f(R)$ gravity~\cite{Eling:2006aw, Akbar:2006mq}.

\section{Logarithmic coefficient in a generic flat FLRW background} \label{app:logCoeffFLRW}

The logarithmic contribution to the von Neumann entropy follows from the heat kernel expansion of the one-loop effective action. For $n_\varphi$ scalars, $n_\psi$ Dirac fermions, and $n_V$ vector fields, 
\begin{equation}
    W = \frac{1}{2} n_\varphi \log \left( \det \Delta_\varphi^{(\xi=1/6)} \right)
    - \frac{1}{2} n_\psi \log ( \det \Delta_\psi )
    + \frac{1}{2} n_V \left[ \log (\det \Delta_V ) - 2 \log \left( \det \Delta_\varphi^{(\xi=0)} \right) \right] \, ,
    \label{eq:eff-action}
\end{equation}
where the last term includes the vector ghosts. The proper-time representation with cutoff $\delta$ is
\begin{equation}
    \log ( \det \Delta )
    = - \int_{\delta^2}^{\infty} \frac{ds}{s} {\rm Tr} \, e^{-s\Delta}
    = - \int d^4 x \sqrt{g}
    \left[ \frac{b_0}{2\delta^4}
    + \frac{b_2}{\delta^2}
    - b_4 \log ( \delta^2 )
    + \text{finite}
    \right] \, ,
    \label{eq:eff-act-exp}
\end{equation}
where the Seeley--DeWitt coefficient $b_4$ controls the logarithmic term~\cite{Vassilevich:2003xt,Eling:2013aqa}. In the presence of a conical singularity, the curvature tensors acquire localized contributions on $\Sigma$, and the bulk integral reduces to a surface term linear in $(1-n)$ as
\begin{equation}
    W(n) = nW_{\rm bulk} + (1-n) W_\Sigma + \mathcal{O} [(1-n)^2 ] \, .
    \label{eq:Wreplica}
\end{equation}
The kinetic operators are
$\Delta \varphi = (-\Box +\xi R)\varphi$ with $\xi = 1/6$, $\Delta\psi = (-\square + R/4)\psi$, and $\Delta V_a = - \square V_a + R_{ab}V^b$.
The surface logarithmic coefficients are~\cite{Eling:2013aqa}
\begin{subequations}
\begin{align}
    W_\Sigma^{\log,\varphi}
    &=
    \frac{1}{360\pi}
    \int_\Sigma d^2y\sqrt{\gamma}
    \left[
    P^{ac}P^{bd}R_{abcd}
    -\frac{1}{2}P^{ab}R_{ab}
    +\frac{5}{2}(1-6\xi)^2R
    \right] \, ,\\
    W_\Sigma^{\log,\psi}
    &=
    \frac{1}{360\pi}
    \int_\Sigma d^2y\sqrt{\gamma}
    \left[
    \frac{7}{2}P^{ac}P^{bd}R_{abcd}
    +2P^{ab}R_{ab}
    -\frac{5}{2}R
    \right] \, ,\\
    W_\Sigma^{\log,V}
    &=
    \frac{1}{360\pi}
    \int_\Sigma d^2y\sqrt{\gamma}
    \left[
    -13P^{ac}P^{bd}R_{abcd}
    +44P^{ab}R_{ab}
    -25R
    \right] \, .
\end{align}
\label{eq:b4_cone}%
\end{subequations}
Here, $P^{ab}$ projects onto the 2D orthogonal space of $\Sigma$. 

For the flat FLRW metric, the nonvanishing components of the Riemann tensor are
\begin{equation}
    R_{0i0j} = -a^2 ( \dot H + H^2) \delta_{ij}  
    \quad \text{and} \quad
    R_{ijkl} = a^4 H^2 ( \delta_{ik}\delta_{jl}-\delta_{il} \delta_{jk} ) \,,
\end{equation}
from which the nonvanishing components of the Ricci tensor follow
\begin{equation}
    R_{00} = -3 (\dot H + H^2 ) 
    \quad \text{and} \quad
    R_{ij} = a^2 ( \dot H + 3 H^2 ) \delta_{ij} \, ,
\end{equation}
giving the Ricci scalar as
\begin{equation}
    R = 6 ( \dot H + 2 H^2 ) \, .
\end{equation}
The components of the normal projector are
\begin{equation}
    P^{00}=-1,
    \quad
    P^{0i}=0,
    \quad \text{and} \quad
    P^{ij}=\frac{x^ix^j}{a^2r^2} \, .
    \label{eq:projectorFLRW}
\end{equation}
At the apparent horizon, $a^2r_A^2=H^{-2}$, so
\begin{equation}
    P^{ac}P^{bd}R_{abcd}=2(\dot H+H^2)
    \quad \text{and} \quad
    P^{ab}R_{ab}=4\dot H+6H^2 \, .
    \label{eq:projectedCurvature}
\end{equation}
Substitution into Eqs.~\eqref{eq:b4_cone} gives
\begin{subequations}
\begin{align}
    W_\Sigma^{\log,\varphi}
    &=
    \frac{1}{90H^2}
    \left[
    2(\dot H+H^2)
    -\frac{1}{2}(4\dot H+6H^2)
    \right]
    =-\frac{1}{90} \, ,\\
    W_\Sigma^{\log,\psi}
    &=
    \frac{1}{90H^2}
    \left[
    7(\dot H+H^2)
    +2(4\dot H+6H^2)
    -15(\dot H+2H^2)
    \right]
    =-\frac{11}{90} \, ,\\
    W_\Sigma^{\log,V}
    &=
    \frac{1}{90H^2}
    \left[
    -26(\dot H+H^2)
    +44(4\dot H+6H^2)
    -150(\dot H+2H^2)
    \right]
    =-\frac{62}{90} \, .
\end{align}
\label{eq:WlogFLRW}%
\end{subequations}
In the trace anomaly of Eq.~\eqref{eq:TrAnomal}, the Euler density, Weyl squared and $\square R$ terms form the standard basis, known as type-A, type-B and type-D, respectively~\cite{Deser:1993yx,Boulanger:2007ab,Awad:2000ac}. $W_\Sigma$ determines the coefficient $\alpha$ of the type-A term in Eq.~\eqref{eq:TrAnomal}, while the type-B term vanishes identically in the spatially flat FLRW.

All $\dot H$ terms in Eq.~\eqref{eq:WlogFLRW} cancel, leaving only the exact dS values~\cite{Casini:2011kv, Eling:2013aqa}. This cancellation is the surface manifestation of the type-A anomaly. We note that for a gauge field, there is an ambiguity regarding the treatment of the boundary modes~\cite{Donnelly:2014fua,Donnelly:2015hxa}. In particular, the geometric replica entropy contains an additional contribution associated with electromagnetic edge modes on the entangling surface $\Sigma$, shifting $W_\Sigma^{\log,V}$ by $-1/3$ relative to the one obtained from mutual information~\cite{Casini:2015dsg,Casini:2019nmu}. Since we use the geometric replica construction, we adopt the edge-inclusive coefficient of $-62/90$. In addition, the contact term for nonminimally coupled scalars does not contribute to the entropy at the logarithmic order for the conformal case~\cite{Solodukhin:1995ak, Donnelly:2015hxa}.

We comment on the validity of Eqs.~\eqref{eq:b4_cone} in our case. For an entangling surface of a physical radius $\mathcal R_\Sigma = a(t) r_\Sigma$, the replica geometry is, in general, a squashed cone, and the surface logarithmic coefficients acquire extrinsic curvature corrections~\cite{Solodukhin:2008dh,Fursaev:2013fta}. These corrections are incorporated into the surface expressions through the replacements 
\begin{equation}\label{eq:PR-shifted}
    P^{ac}P^{bd}R_{abcd}\rightarrow P^{ac}P^{bd}R_{abcd}-k^i_{ab}k_i^{ab} 
    \quad \text{and} \quad
    P^{ab}R_{ab}\rightarrow P^{ab}R_{ab}-\frac{1}{2} k_i k^i \, ,
\end{equation}
where $k_i=\gamma^{ab}k_{iab}$. In the spatially flat FLRW metric, one obtains 
\begin{equation} 
\label{eq:extcurv_sq}
    k_i k^i=4(\mathcal R_\Sigma^{-2}-H^2) 
    \quad \text{and} \quad
    k^i_{ab}k_i^{ab} = 2 (\mathcal R_\Sigma^{-2}-H^2) \, . 
\end{equation}
Here, the normal indices are contracted with the Lorentzian normal metric $\eta_{ij}=\operatorname{diag}(-1,1)$. Both quadratic invariants therefore vanish at the apparent horizon, $\mathcal R_\Sigma = H^{-1} $. Consequently, the extrinsic curvature corrections vanish on the surface considered here, and Eqs.~\eqref{eq:b4_cone} apply without modification.

The spacetime independence of the type-A anomaly coefficient $\alpha$ can also be understood from the Wess--Zumino (WZ) consistency condition. Through $\alpha=90W_\Sigma^{\rm log}$ in Eq.~\eqref{eq:deltaS-Tranomal}, this is consistent with the constant surface coefficient obtained above. We stress, however, that the absence of additional dependence on $H$ and its derivatives at the FLRW apparent horizon follows from the explicit cancellation in Eq.~\eqref{eq:WlogFLRW}, rather than from WZ consistency alone.

To see directly why the type-A coefficient cannot depend on spacetime position, suppose provisionally that $\alpha=\alpha(x)$, while retaining this anomaly basis and introducing no additional terms involving derivatives of $\alpha$. Since local Weyl transformations commute, WZ consistency requires~\cite{Wess:1971yu}
\begin{equation}
    [\delta_{\sigma_1},\delta_{\sigma_2}]
    \Gamma_X[g]=0 \, .
    \label{eq:WZcondition}
\end{equation}
Using Eq.~\eqref{eq:dW-Tanomal} and
\begin{equation}
    \delta_\sigma (\sqrt{-g}\,E_4)
    =
    8\sqrt{-g}\,G^{ab}\nabla_a\nabla_b\sigma \, ,
\end{equation}
the type-A contribution to the commutator becomes
\begin{equation}
    0 =
    \left. [\delta_{\sigma_1},\delta_{\sigma_2}]
    \Gamma_X[g]\right|_{\rm A} = 
    \frac{1}{720\pi^2}
    \int d^4x\sqrt{-g}\,
    \alpha(x)G^{ab}
    \left(
    \sigma_2\nabla_a\nabla_b\sigma_1
    -
    \sigma_1\nabla_a\nabla_b\sigma_2
    \right) \, .
    \label{eq:WZ-var-action}
\end{equation}
Integrating by parts and using $\nabla_aG^{ab}=0$ gives
\begin{equation}
    0
    =
    \int d^4x\sqrt{-g}\,
    G^{ab}\nabla_a\alpha
    \left(
    \sigma_1\nabla_b\sigma_2
    -
    \sigma_2\nabla_b\sigma_1
    \right) \, .
    \label{eq:WZ-alpha-gradient}
\end{equation}
Since this relation must hold for arbitrary local Weyl parameters $\sigma_1$ and $\sigma_2$ and for arbitrary background metrics, it follows that
\begin{equation}
    \nabla_a\alpha=0 \, .
\end{equation}
Thus, $\alpha$ is a spacetime-independent coefficient fixed by the conformal field content.

We also verify the simplification of Eq.~\eqref{eq:fin-Weyl-entropy}. Among the three contributions of the 4D Weyl anomaly, the surface integral after the local $R^2$ entropy contribution is separated off as in Sec.~\ref{sec:EE-conf} is governed by the type-A and type-B anomalies~\cite{Solodukhin:2011gn,Solodukhin:2013yha}. The type-A contribution of $E_4$ reduces to the intrinsic curvature $R_\Sigma$ of the entangling surface, while the type-B term induces 
\begin{equation}
    K_\Sigma
    = P^{ac}P^{bd}R_{abcd} - P^{ab}R_{ab} + \frac{1}{3}R - \left( k^i_{ab}k_i^{ab} - \frac{1}{2}k_ik^i \right)
    \label{eq:KSigma-def}
\end{equation}
at the surface integral. Under a Weyl transformation~\cite{Solodukhin:2008dh,Solodukhin:2013yha},
\begin{equation}
    R_\Sigma(e^{2\sigma}\gamma)
    =e^{-2\sigma} \left[ R_\Sigma(\gamma) - 2 \Delta_\Sigma \sigma \right]
    \quad \text{and} \quad
    K_\Sigma(e^{2\sigma}g,e^{2\sigma}\gamma)
    = e^{-2\sigma} K_\Sigma(g,\gamma) \, .
    \label{eq:Curv-transf}
\end{equation}
For $\sigma=\log a(t)$, the factor is constant on a fixed-time sphere, so
\begin{equation}
    D_\Sigma \sigma = 0 
    \quad \text{and} \quad
    \Delta_\Sigma \sigma=0 \, ,
    \label{eq:Dsigma0}
\end{equation}
which gives
\begin{equation}
    \sqrt{\gamma'} R_\Sigma(\gamma') = \sqrt{\gamma} R_\Sigma(\gamma) \, .
\end{equation}
Moreover, the Weyl contribution and the traceless extrinsic curvature contribution vanish separately as in Eqs.~\eqref{eq:projectedCurvature} and \eqref{eq:extcurv_sq},
\begin{equation}
    P^{ac}P^{bd}R_{abcd} -P^{ab}R_{ab} + \frac{1}{3}R = 0 
    \quad \text{and} \quad
    k^i_{ab} k_i^{ab} = k_i k^i = 0 \, ,
    \label{eq:Curv-vanish}
\end{equation}
so that $K_\Sigma=0$. Together with $\int_\Sigma\sqrt{\gamma} \, R_\Sigma=8\pi$, this reduces Eq.~\eqref{eq:fin-Weyl-entropy} to Eq.~\eqref{eq:SvNWeylTr} with no additional dependence on $H$ or its derivatives.

\section{Canonical density matrix and the renormalized von Neumann entropy} \label{app:denmat}

We give an independent canonical derivation of Eq.~\eqref{eq:dSvNdt}, using the conformally coupled scalar as an example. In conformal coordinates, the metric is $ds^2 = a^2(\eta) ( -d\eta^2 + \delta_{ij} dx^i dx^j )$ and we define $\mathcal H \equiv a' / a$ and $\tilde \varphi \equiv a \varphi$, where a prime denotes differentiation with respect to conformal time $\eta$. For a massless conformally coupled scalar, the action becomes
\begin{equation}
    S_\varphi = \frac{1}{2} \int d \eta \, d^3 x \left[ (\tilde\varphi')^2 - ({\nabla} \tilde\varphi)^2 \right] - \frac{1}{2} \left[ \mathcal H (\eta) \int d^3x \, \tilde\varphi^2 \right]_{\eta_i}^{\eta_f} \, .
    \label{eq:Sconf}
\end{equation}
The bulk action is therefore the Minkowski space action for $\tilde\varphi$, while the difference is the temporal boundary term. For the conformal vacuum, the initial wave functional may be chosen consistently with this boundary term. The Schr\"odinger wave functional at time $\eta$ can then be written as
\begin{equation}
    \Psi_{\rm FLRW}[\varphi,\eta] = N_a(\eta) \exp \left[ - \frac{i}{2} \mathcal H (\eta) \int d^3x \, \tilde\varphi^2 \right] \Psi_{\rm Mink}[\tilde\varphi, \eta] \, ,
    \label{eq:wavefunctional}
\end{equation}
where $\Psi_{\rm Mink}$ is the Minkowski vacuum wave functional and $N_a$ is the normalization associated with the field rescaling.

The relation between the two sets of canonical variables is particularly useful for the reduced density matrix. With $\pi_\varphi=a^2\varphi'$ and $\pi_{\tilde\varphi} = \tilde\varphi'$, one finds
\begin{equation}
    \begin{pmatrix}
        \varphi\\
        \pi_\varphi
    \end{pmatrix}
    =
    \begin{pmatrix}
        a^{-1}&0\\
        -a'&a
    \end{pmatrix}
    \begin{pmatrix}
        \tilde\varphi\\
        \pi_{\tilde\varphi}
    \end{pmatrix} \, .
    \label{eq:canonicalMap}
\end{equation}
The transformation has unit determinant and preserves the symplectic form. On a regulated spatial lattice, it is implemented pointwise so that both the field rescaling and the phase in Eq.~\eqref{eq:wavefunctional} act independently at each lattice site. The corresponding unitary operator therefore factorizes with respect to any spatial bipartition.

Let $B$ be a ball of comoving radius $r$ at fixed conformal time and $\bar B$ its complement. With the spatial regulator in place, the Hilbert space factorizes as
\begin{equation}
    \mathcal H_{\rm reg} = \mathcal H_B \otimes \mathcal H_{\bar B} \, .
\end{equation}
The locality of the transformation can also be seen directly from
\begin{equation}
    \int d^3x\,\tilde\varphi^2 = \int_B d^3 x \, \tilde\varphi^2 + \int_{\bar B} d^3 x \, \tilde\varphi^2 \, .
    \label{eq:appPhaseSplit}
\end{equation}
Therefore, at each fixed time, the unitary operator that implements the complete canonical transformation takes the form
\begin{equation}
    W(\eta) = W_B(\eta) \otimes W_{\bar B}(\eta) \, ,
\end{equation}
and
\begin{equation}
    |\Psi_{\rm FLRW}(\eta)\rangle = \left[ W_B(\eta) \otimes W_{\bar B}(\eta) \right] |\Psi_{\rm Mink}\rangle \, .
    \label{eq:stateLocalUnitary}
\end{equation}
For the corresponding density matrix, taking the partial trace over $\bar B$ gives
\begin{equation}
    \rho_B^{\rm FLRW}(\eta;r) = W_B(\eta) \rho_B^{\rm Mink}(r) W_B^\dagger(\eta) \, .
    \label{eq:localunitary}
\end{equation}
Consequently,
\begin{equation}
    {\rm Spec} \, \rho_B^{\rm FLRW}(\eta;r) = {\rm Spec} \, \rho_B^{\rm Mink}(r)
    \label{eq:spectrum}
\end{equation}
and hence
\begin{equation}
    S_n^{\rm FLRW}(\eta;r) = S_n^{\rm Mink}(r) 
    \quad \text{and} \quad 
    S_{\rm vN}^{\rm FLRW}(\eta;r) = S_{\rm vN}^{\rm Mink}(r)
    \label{eq:entropyEqualityReg}
\end{equation}
for the regulated theory. Thus, the equality extends to the complete regulated entanglement spectrum, rather than only to its universal logarithmic coefficient.

To apply this result to the apparent horizon, the region size and the UV regulator must be compared in the same physical prescription. We substitute $r_B = 1/ [a(\eta)H(\eta)]$ while a fixed physical short-distance cutoff $\delta_{\rm phys}$ corresponds to the comoving cutoff $\delta(\eta) = \delta_{\rm phys}/a(\eta)$. Therefore, the ratio $r_B(\eta) / \delta(\eta) = 1 / [ H(\eta)\delta_{\rm phys}]$ carries no additional dependence on the scale factor, beyond that contained in the physical radius of the apparent horizon. 

The local unitary argument above is, by construction, a statement about the regulated theory. It fixes the dependence of the regulated field entropy on the dimensionless ratio $r_B/\delta$, but does not by itself specify how the local UV divergences are assigned between the renormalized field entropy and the gravitational couplings. In particular, for the apparent horizon region and the fixed physical cutoff introduced above, the universal logarithmic contribution to the entropy associated with a spherical entangling surface is
\begin{equation}
    S_{\rm vN}^{\rm reg} \supset \frac{\alpha}{90} \log \left(\frac{r_B}{\delta} \right) 
    = 
    \frac{\alpha}{90} \log \left( \frac{1}{H\delta_{\rm phys}} \right) \, . \label{eq:SvNregulatedCanonical}
\end{equation}
Equation~\eqref{eq:SvNregulatedCanonical} is the regulated expression underlying the renormalized $H$ dependence used in Sec.~\ref{sec:EE-conf}. 

To pass to the renormalized field entropy, we use the same assignment of local UV terms to the gravitational couplings. The power-law divergence is absorbed into the renormalized Newton's constant, while the local logarithmic divergence is absorbed into the renormalized curvature-squared couplings. After these local terms have been assigned consistently to the renormalized gravitational action, the remaining radius dependence of the field entropy can be written as
\begin{equation}
    S_{\rm vN}^{\rm rn} = \frac{\alpha}{90} \log \left[ \frac{\mu}{H(\eta)} \right] + C_0 \, ,
    \label{eq:SvNcanonical}
\end{equation}
where $\mu$ is a fixed renormalization scale and $C_0$ is independent of the radius of the spherical surface for a fixed renormalization prescription. A change of the renormalization scale $\mu$ shifts $C_0$ by a radius-independent constant and therefore does not affect the entropy
rate.

Since neither $\mu$ nor $C_0$ depends on time, Eq.~\eqref{eq:SvNcanonical} gives
\begin{equation}
    {\dot S}_{\rm vN}^{\rm rn} = - \frac{\alpha}{90} \frac{\dot H}{H} = \frac{\alpha}{90}H\epsilon \, ,
\end{equation}
which is Eq.~\eqref{eq:dSvNdt}. The canonical derivation therefore shows that, for the conformal vacuum, there is no additional state-dependent time dependence of the field entropy beyond the dependence on the apparent horizon radius.

The above derivation relies on conformal invariance. For a massive scalar or nonconformal curvature coupling, the action contains the bulk term
\begin{equation}
    S_\varphi \supset - \frac{1}{2} \int d \eta \, d^3 x \left[ a^2m^2 + (6 \xi - 1) \frac{a''}{a} \right] \tilde\varphi^2 \, .
    \label{eq:nonconformal}
\end{equation}
This term is part of the bulk dynamics and cannot be absorbed into the temporal boundary term in Eq.~\eqref{eq:Sconf}. Consequently, the pointwise canonical transformation used above no longer reduces the FLRW problem to the Minkowski vacuum problem, and Eqs.~\eqref{eq:localunitary}--\eqref{eq:entropyEqualityReg} do not follow. In such theories, the entanglement spectrum can acquire additional state-dependent time dependences.

\section{Exact $R^2$ identity and Einstein-frame description}
\label{app:Einsteinframe}

We first derive an exact relation between the heat flow and Wald entropy contributions of the local $R^2$ term. We then express the full $f(R)$ apparent horizon balance in the conformally related Einstein-frame description. No expansion in the Hubble-flow parameters is used unless stated otherwise.

Let $\delta Q_{R^2}$ denote the inward heat flow associated with the effective stress tensor obtained from the local $C_{R^2}R^2$ term, with the Einstein--Hilbert term kept on the gravitational side of the field equation. Either of the two consistently paired apparent horizon prescriptions discussed in App.~\ref{app:tempConf} may be used, since they give the same ratio $\delta Q/T$.

It is convenient to write the local gravitational Lagrangian as
\begin{equation}
    \mathcal{L} \supset \frac{1}{16\pi G} f(R) \, , 
\end{equation}
where
\begin{equation}
    f(R)=R+16\pi G C_{R^2}R^2 \, ,
\end{equation}
and define
\begin{equation}
    F\equiv \frac{df}{dR} = 1+32\pi G C_{R^2}R \, . \label{eq:FdefApp}
\end{equation}
When the contribution of the $R^2$ term is written as an effective stress tensor, its energy density and pressure in the spatially flat FLRW metric satisfy
\begin{equation}
    8\pi G ( \rho_{R^2}+p_{R^2} ) = \ddot F - H \dot F + 2(F-1) \dot H \, .
    \label{eq:R2rhoPApp}
\end{equation}
Using Eq.~\eqref{eq:entropyflow}, we obtain
\begin{equation}
    \frac{1}{T} \frac{\delta Q_{R^2}}{dt} = \frac{\pi}{GH^3} \left[ H \dot F - 2 (F-1) \dot H - \ddot F \right] \, .
    \label{eq:R2HeatExactApp}
\end{equation}

The contribution of the same local $R^2$ term to the Wald entropy is
\begin{equation}
    S_{W,R^2} = \frac{(F-1)A}{4G} = \frac{\pi(F-1)}{GH^2} \, , \label{eq:R2WaldEntropyApp}
\end{equation}
where $A = 4\pi H^{-2}$. Its time derivative is therefore
\begin{equation}
    {\dot S}_{W,R^2} = \frac{\pi}{GH^3} \left[ H \dot F - 2(F-1) \dot H \right] \, .  \label{eq:R2WaldRateApp}
\end{equation}
Subtracting Eq.~\eqref{eq:R2HeatExactApp} from Eq.~\eqref{eq:R2WaldRateApp} gives
\begin{equation}
    {\dot S}_{W,R^2} - \frac{1}{T} \frac{\delta Q_{R^2}}{dt} = \frac{d_i S}{dt} = \frac{\pi}{GH^3} \ddot F \, , \label{eq:diSexact}
\end{equation}
which is Eq.~\eqref{eq:diSdtCompact}. Expanding Eq.~\eqref{eq:diSexact} in the Hubble-flow parameters reproduces Eq.~\eqref{eq:diSdt} and, at the first nonvanishing order, Eq.~\eqref{eq:diSleading}. The right-hand side of Eq.~\eqref{eq:diSexact} is not of definite sign, independently of any Hubble-flow expansion.

We now describe the same relation in the Einstein-frame. Consider the Jordan-frame $f(R)$ theory in a spatially flat FLRW metric, with expansion sourced by a homogeneous, minimally coupled inflaton $\phi$. For $F>0$, we introduce the Einstein-frame metric $g_{ab}^E$ and the scalaron $\chi$ respectively as
\begin{align}
    g_{ab}^E & = Fg_{ab} \, , \\
    \chi & = \sqrt{\frac{3}{2}} m_{\rm Pl} \log F \, , 
    \label{eq:frameMapMain}
\end{align}
where $m_{\rm Pl}=1/\sqrt{8\pi G}$. The cosmic time and scale factor in the Einstein-frame are given by 
\begin{equation}
    dt_E = \sqrt F \, dt 
    \quad \text{and} \quad 
    a_E = \sqrt F \, a \, ,
    \label{eq:frameTimeScaleApp}
\end{equation}
and hence
\begin{equation}
    H_E = \frac{H}{\sqrt F}(1+u) \, , 
    \label{eq:HEMain}
\end{equation}
where $u \equiv F/(2FH)$. Throughout this appendix, an overdot denotes differentiation with respect to the Jordan-frame cosmic time $t$.

For a Jordan-frame two-sphere $\Sigma_J$ and its conformal image $\Sigma_E^{\rm map}$ in the Einstein-frame, the Jordan-frame Wald entropy is
\begin{equation}
    S_W^J[\Sigma_J] = \frac{F A_J}{4G} = \frac{A_E^{\rm map}}{4G} \, .
    \label{eq:sameSurfaceEntropyMain}
\end{equation}
Thus, the Jordan-frame Wald entropy and the Einstein-frame Bekenstein--Hawking entropy agree when they are evaluated on surfaces related by the conformal transformation~\cite{Jacobson:1993vj}.

The apparent horizon defined directly by the Einstein-frame metric is, however, not in general the conformal image of the Jordan-frame apparent horizon. In the spatially flat FLRW metric, the Einstein-frame apparent horizon has areal radius $H_E^{-1}$, whereas the conformal image of the Jordan-frame apparent horizon has areal radius $\sqrt F \, H^{-1}$. Their ratio is
\begin{equation}
    \frac{H_E^{-1}}{\sqrt F \, H^{-1}} = (1+u)^{-1} \, .
    \label{eq:horizonRadiusRatioApp}
\end{equation}

For definiteness, we use the isochoric temperatures
\begin{equation}
    T_J = \frac{H}{2\pi} 
    \quad \text{and} \quad
    T_E=\frac{H_E}{2\pi} \, ,
\end{equation}
corresponding to $T_H$ in App.~\ref{app:tempConf}. All heat flows below follow the inward-flow convention. Einstein-frame rates are expressed per unit Jordan-frame time using Eq.~\eqref{eq:frameTimeScaleApp}.

In the Einstein-frame, the gravitational sector has the Einstein--Hilbert form, while $\phi$ and $\chi$ appear as scalar fields. The apparent horizon thus saturates the Clausius inequality without a separate higher-curvature entropy residual~\cite{Cai:2005ra}, hence in our inward-flow convention, 
\begin{equation}
    \left. {\dot S}_{\rm area}^{E,{\rm AH}} \right|_t + \frac{1}{T_E} \frac{\delta Q_\phi^E}{dt} + \frac{1}{T_E} \frac{\delta Q_\chi^E}{dt} = 0 \, .
    \label{eq:EinsteinAreaBalanceApp}
\end{equation}
The corresponding Jordan-frame relation for $f(R)$ gravity with the inflaton stress tensor on the matter side is
\begin{equation}
    {\dot S}_W^J + \frac{1}{T_J} \frac{\delta Q_\phi^J}{dt} = \frac{d_i S}{dt} \, .
    \label{eq:JordanBalanceApp}
\end{equation}
Subtracting Eq.~\eqref{eq:EinsteinAreaBalanceApp} from Eq.~\eqref{eq:JordanBalanceApp} gives
\begin{equation}
    \frac{d_i S}{dt} = \left( {\dot S}_W^J - \left. {\dot S}_{\rm area}^{E,{\rm AH}} \right|_t \right) + \left( \frac{1}{T_J} \frac{\delta Q_\phi^J}{dt} - \frac{1}{T_E} \frac{\delta Q_\phi^E}{dt} \right) - \frac{1}{T_E} \frac{\delta Q_\chi^E}{dt} \, .
    \label{eq:frameResidualMain}
\end{equation}
Equation~\eqref{eq:frameResidualMain} follows from the field equations in the two frames together with the apparent horizon defined by each metric and does not require a Hubble-flow expansion. Equation~\eqref{eq:frameResidualMain} is the Einstein-frame decomposition of Eq.~\eqref{eq:diSexact} or Eq.~\eqref{eq:diSdtCompact}.

The first parenthesis in Eq.~\eqref{eq:frameResidualMain} arises because the two entropy rates are evaluated on different surfaces. Using Eq.~\eqref{eq:sameSurfaceEntropyMain}, it can be written as
\begin{equation}
    {\dot S}_W^J - \left. {\dot S}_{\rm area}^{E,{\rm AH}} \right|_t = \frac{d}{dt} \left( \frac{A_E^{\rm map} - A_E^{\rm AH}}{4G} \right) \, .
    \label{eq:areaDifferenceApp}
\end{equation}
It is therefore the difference between the Einstein-frame area entropy evaluated on the conformal image of the Jordan-frame apparent horizon and on the Einstein-frame apparent horizon itself.

The second parenthesis in Eq.~\eqref{eq:frameResidualMain} is the difference between the inflaton entropy flow evaluated using the Jordan- and Einstein-frame metrics, respectively. It should not be interpreted as an additional energy flux through the region between the two horizon surfaces. For the homogeneous inflaton in the Jordan-frame,
\begin{equation}
    \rho_\phi^J+p_\phi^J = \dot\phi^2 \, ,
\end{equation}
whereas after the conformal transformation,
\begin{equation}
    \rho_\phi^E+p_\phi^E = \frac{1}{F} \left( \frac{d\phi}{dt_E} \right)^2 = \frac{\dot\phi^2}{F^2} \, .
    \label{eq:phiRhoPFramesApp}
\end{equation}
Together with Eqs.~\eqref{eq:frameTimeScaleApp} and  \eqref{eq:HEMain}, this gives
\begin{equation}
    \frac{1}{T_E} \frac{\delta Q_\phi^E}{dt} = (1+u)^{-3} \frac{1}{T_J} \frac{\delta Q_\phi^J}{dt} \, .
    \label{eq:phiHeatTransformApp}
\end{equation}
Thus, the two inflaton entropy flows differ because the scalar stress tensor, cosmic time, apparent horizon radius, and temperature are related nontrivially by the conformal transformation.

The final term in Eq.~\eqref{eq:frameResidualMain} is the Einstein-frame entropy flow of the scalaron. Since the scalaron has a canonical kinetic term in the Einstein-frame,
\begin{align}
    -\frac{1}{T_E} \frac{\delta Q_\chi^E}{dt} &= \frac{8\pi^2}{\sqrt F\,H_E^3}\dot\chi^2 \nonumber\\
    &= \frac{3\pi}{2G} \frac{\dot F^2}{F H^3} (1+u)^{-3} \geq 0 \, , 
    \label{eq:scalaronFluxMain}
\end{align}
where $\dot\chi = d\chi/dt$. On the branch where both frames are expanding, $H>0$ and $H_E>0$, the inequality follows from the null energy condition for a canonical scalar,
\begin{equation}
    T_{ab}^{(\chi)}k^ak^b
    = \left(k^a\nabla_a\chi\right)^2 \geq 0
\end{equation}
for any null vector $k^a$. We emphasize that this does not imply $d_i S / dt$ has a definite sign. See also Ref.~\cite{Chirco:2010sw} for a similar heat flow interpretation of a scalar degree of freedom arising from $f(R)$ gravity.

We finally clarify the power counting in $C_{R^2}$ when comparing Eqs.~\eqref{eq:diSexact} and \eqref{eq:frameResidualMain}. Although the two equations are equivalent, terms quadratic and higher order in $C_{R^2}$ appear explicitly in the individual contributions to Eq.~\eqref{eq:frameResidualMain}, like the scalaron entropy flow in Eq.~\eqref{eq:scalaronFluxMain}.

To isolate the explicit dependence on $C_{R^2}$, we hold the FLRW geometry fixed, treating $H(t)$ and hence $R(t)$ as prescribed functions of time. If instead one varies $C_{R^2}$ as a parameter of the gravitational theory and solves the field equations anew, the background geometry generally changes as well, introducing additional implicit dependence on $C_{R^2}$ through the resulting solution. Such dependence of the background on the gravitational action is not the power counting considered here. At fixed geometry, both $\ddot F$ and the right-hand side of Eq.~\eqref{eq:diSexact} are exactly linear in $C_{R^2}$.

Even at fixed geometry, however, the individual terms in the Einstein-frame decomposition \eqref{eq:frameResidualMain} contain nonlinear functions of $F$, and hence of $C_{R^2}$. For example, Eq.~\eqref{eq:scalaronFluxMain} begins at $\mathcal O(C_{R^2}^2)$ because it is proportional to $\dot F^2$. Only after all the terms in Eq.~\eqref{eq:frameResidualMain} are combined do the quadratic and higher-order contributions cancel, thereby reproducing the exactly linear Jordan-frame expression \eqref{eq:diSexact} for the same FLRW geometry.

\end{widetext}

\bibliography{references}

\end{document}